\documentclass[aps,prb,longbibliography]{revtex4-2}

\usepackage{graphicx}
\usepackage{upgreek}%
\usepackage{cancel}
\usepackage{dcolumn}
\usepackage{bm}
\usepackage{xcolor}%
\usepackage{hyperref}

\usepackage{latexsym}
\usepackage{etoolbox}

\usepackage[T1]{fontenc}             
\usepackage[utf8]{inputenc}  
\usepackage[english]{babel}
\usepackage[babel]{csquotes}
\usepackage{newlfont}
\usepackage{color}
\usepackage{booktabs}
\usepackage{verbatim}
\usepackage{braket}
\usepackage{graphicx}
\usepackage{float}
\usepackage{siunitx}
\usepackage{mathtools}
\allowdisplaybreaks
\usepackage{nicefrac}
\let\oldAA\AA
\renewcommand{\AA}{\text{\normalfont\oldAA}}

\newcommand{\editor}[2]{%
  \expandafter\newcommand\csname #1note\endcsname[1]{%
    \textcolor{#2}{(\textbf{#1:} \textit{##1})}}%
  \expandafter\newcommand\csname #1\endcsname[1]{%
    \textcolor{#2}{##1}}%
  \expandafter\newcommand\csname #1cancel\endcsname[1]{%
    \textcolor{#2}{\sout{##1}}}%
  \expandafter\newcommand\csname #1change\endcsname[2]{%
    \textcolor{#2}{\sout{##1} ##2}}%
  \newenvironment{#1text}{\color{#2}}{\color{black}}
}

\definecolor{brown}{rgb}{0.633,0.156,0.156}
\definecolor{pink}{rgb}{1,0,1}
\editor{AG}{red}
\editor{CK}{pink}
\editor{PB}{blue}

\newcommand{\qb}{q_{\mathrm{3D}}}
\newcommand{\qm}{q}
\newcommand{\qmv}{\mathbf{q}}
\newcommand{\qz}{q_z}
\newcommand{\crsec}{S}
\newcommand{\kfact}{A}

\begin{document}

\title{Direct observation of the vanishing electron energy-loss cross section in graphene}

\author{Alberto Guandalini}
\affiliation{Dipartimento di Fisica, Universit\`a di Roma La Sapienza, Piazzale Aldo Moro 5, I-00185 Roma, Italy}
\email{alberto.guandalini@uniroma1.it}

\author{Ryosuke Senga}
\affiliation{AIST Tsukuba, 1-1-1 Umezono, Tsukuba, Ibaraki 305-8560 Japan}
	
\author{Yung-Chang Lin}
\affiliation{AIST Tsukuba, 1-1-1 Umezono, Tsukuba, Ibaraki 305-8560 Japan}
	
\author{Kazu Suenaga}
\affiliation{AIST Tsukuba, 1-1-1 Umezono, Tsukuba, Ibaraki 305-8560 Japan}
\affiliation{The Institute of Scientific and Industrial Research (SANKEN), Osaka University Mihogaoka 8-1, Ibaraki, Osaka 567-0047, Japan }
	
\author{Paolo Barone}
\affiliation{CNR-SPIN, Area della Ricerca di Tor Vergata, Via del Fosso del Cavaliere 100, I-00133 Rome, Italy}
\affiliation{Dipartimento di Fisica, Universit\`a di Roma La Sapienza, Piazzale Aldo Moro 5, I-00185 Roma, Italy}
	
\author{Francesco Mauri}
\affiliation{Dipartimento di Fisica, Universit\`a di Roma La Sapienza, Piazzale Aldo Moro 5, I-00185 Roma, Italy}
		
\author{Thomas Pichler}
\affiliation{University of Vienna, Faculty of Physics, Strudlhofgasse 4, A1090, Austria}

\author{Christian Kramberger}
\affiliation{The Institute of Scientific and Industrial Research (SANKEN), Osaka University Mihogaoka 8-1, Ibaraki, Osaka 567-0047, Japan }
\affiliation{University of Vienna, Faculty of Physics, Strudlhofgasse 4, A1090, Austria}


\begin{abstract}
In transmission electron energy-loss spectroscopy, the  cross section in 2D is quenched by kinematic effects once the momentum transfer becomes smaller than a critical value set by $q_z$, the momentum loss parallel to the beam.
Our highly momentum ($\Delta q = 0.02$~\AA$^{-1}$) and energy ($\Delta E = 45$~meV) resolved setup is instrumental on delivering the unprecedented experimental verification of quenched 2D EEL spectra on freestanding graphene at momentum transfers $q$ below $0.06$~\AA$^{-1}$.
We retrieve the intrinsic uniform dielectric response of graphene from measured spectra by quantifying the kinematic suppression.
\end{abstract}
\maketitle

Electron energy-loss spectroscopy (EELS) is a powerful tool to investigate electronic excitations in materials, as it directly probes the longitudinal dielectric response of the system, in principle resolved both in energy and momentum \cite{Egerton_book,Ibach_book}. 
Special attention is often devoted to vanishing momentum transfer, also referred to as the optical limit, where the cross section contains the same information delivered by the optical response of the sample~\cite{Mondio_92,Hong_21}.
Energy resolution of few meV can be attained, e.g., by using
low-energy electrons that are reflected by the sample surface, the scattering angle enabling the measure of  the momentum transfer and thus providing access to the excitation\rq s dispersion with a momentum resolution that has been recently pushed down to 10$^{-3}$\AA$^{-1}$\cite{jiandong2015_2dhreels.res}.
This technique, known as high-resolution EELS (HREELS), is however a surface-sensitive one, and as such it
cannot probe the bulk response of the target material, since the scattered electrons hardly penetrate it. On the contrary, higher energy electrons as those used in a transmission electron microscope (TEM) can pass through a target of about 1~$\mu$m thickness \cite{Egerton_2009}.
A sketch of the experimental geometry is shown in the left panel of Fig.~\ref{figIntro}. If the thickness $d$ of the sample is such that $\qb d \gg 1$, where $\qb = |\mathbf{k}_i-\mathbf{k}_f|$ is the momentum lost by the beam electron,
the loss function can be modelled within inelastic scattering theory by approximating the dielectric response of the film with that of the bulk~\cite{Sturm_1993,Egerton_2009,Keast_2016}. 
Otherwise, finite thickness effects become relevant. 
Recent years have witnessed a significant effort in optimizing both energy and momentum resolution of TEM-EELS, reaching few tens of meV and few hundredths of \AA$^{-1}$, respectively \cite{Senga_2019,Li2020,Hage2020,Senga2022,Hage_2018,Hong_2020,guandalini_nanolett2023}. Such advancement of the experimental setup has made TEM-EELS a valuable tool for studying bulk fundamental excitations as a function of momentum transfer in small volumes of material \cite{Pichler1998,Neudert_prl1998,Schuster_prl2007,Kramberger2008,Schuster_2009}. 

The angular and energy dependence of the inelastic electron scattering are specified by a double-differential cross section, that can be generally expressed as the product of two terms: the so-called loss function, accounting for the target response to an incident electron, and a kinematic prefactor that instead only depends on the scattering geometry through energy and momentum conservation~\cite{Egerton_book,Ibach_book}.
 The kinematic prefactor, crucial in the interpretation of HREELS spectra \cite{Mills_1975, Lambin_1987, Palmer_prl1987,Hogan_2009,Li_surfsci2022} because of the dependence of the momentum transfer on both the scattering angle and the energy loss~\cite{Egerton_book}, is generally simplified in TEM-EELS, where the momentum transfer is regarded to lie only within the plane perpendicular to the incoming electron beam \cite{Roth_jes2014}. This assumption is reasonable in typical experimental setups with small scattering angles and momentum resolution of few tenths of \AA$^{-1}$ limiting the access to the low-momentum transfer regime, and as long as the targeted film is thick enough to be viewed as a bulk 3D sample. The TEM-EELS cross section is thus proportional to $-\mathrm{Im}[\varepsilon^{-1}(\qb,\omega)]/\qb^2$, with the energy loss $\hbar\omega$, where the loss function is given by the bulk 3D inverse dielectric function $\varepsilon^{-1}$.

In the last decades, particular attention has been devoted to the study of 2D materials, i.e., materials that are periodic in two directions ($x,y$) and atomically thin in the third direction ($z$), due to their remarkable electronic and optical properties~\cite{Ferrari_2015,Bhimanapati_2015}.
Their intrinsic dielectric response, that in HREELS measurements is affected by the inevitable presence of substrates, can be naturally accessed by TEM-EELS enabling the study of electronic~\cite{Hong_2020,Woo_2023} and lattice~\cite{Hage_2018,Senga_2019} excitations of freestanding 2D materials.
When the sample thickness becomes atomic and/or the system is probed near to the optical limit, dimensionality effects may arise in the dielectric response due to the partially confined nature of the electrons.
In this regime, the system response becomes anisotropic, and one conveniently decomposes the momentum lost by the incoming electrons $\bm q_{3D}$ in  $\qmv$ and $q_z$, the momentum transfer parallel and perpendicular, respectively, to the $xy$ plane of the system [see (a) panel of Fig.~\ref{figIntro}].
While $\qmv$ is transferred to the crystal quasi-momentum of the 2D periodic lattice, thus being a good quantum number for the material's electrons, $\qz$ is not. The latter encodes the dependence of the momentum transfer on the energy loss, while the intrinsic response of the system depends mainly on $\qmv$~\cite{Egerton_book,Roth_jes2014}.
We point out that these considerations, relevant for 2D materials, apply also to thin films, as the interpretation of EEL spectra in terms of the 3D response is an approximation valid only for $\qb d \gg 1$.

Transmission EELS has been studied at different levels of theory mostly within a 3D framework~\cite{Egerton_book, Keast_2016, Nicholls_2021}.
Nonetheless, 
the relation between the measured EELS cross section and the dielectric response of a 2D system has been recently addressed by Nazarov~\cite{Nazarov_2015}, showing that it indeed differs from the 3D response.
As an example, the kinematic term for the case of a 2D system at low momentum transfers significantly differs from the bulk counterpart, due to the $\approx e^{-\qm z}$ decay of electric fields in the vacuum.
In this regime, the 2D material may be approximated as a 2D sheet~\cite{Sohier2015}, displaying qualitatively different dielectric properties~\cite{Cudazzo_2011}.
Remarkably, it has been theoretically predicted that the cross section of a 2D material vanishes in the low-momentum transfer regime instead of giving a maximum signal as the $1/\qb^2$ prefactor would predict, due to the inability of 2D materials to screen the Coulomb interaction in the long-wavelength limit~\cite{Nazarov_2015}.
A TEM setup allowing for a very high momentum resolution is instrumental to probe the 2D features of the kinematic response, that we anticipate occur at $\qm \le \qz$.

In this work, we study the EEL cross section of a prototypical 2D material, graphene, in the optical limit regime, i.e.\ at
low-momentum transfers ($\qm < 0.1$ ~\AA$^{-1}$).
Graphene is one of the most studied 2D materials due to its potential electronic applications, e.g., for plasmonic and electronic devices~\cite{CastroNeto_rev2009,Bonaccorso_2010,Grigorenko_2012, Garcia_2014}. The energy loss function of undoped graphene and of related sp$^2$ carbon materials, graphite and nanotubes, always displays $\pi$ and $\pi+\sigma$ plasmons, arising from interband transitions occurring below $10$ eV and above $15$ eV, respectively. 
Our graphene sample is suspended on a TEM grid and described in detail in the supporting information~\cite{supp-info}.
We consider spurious doping effects, possibly originating from the sample preparation, to be negligible, due to the lack of the Drude-plasmon contribution in the EELS.
The same sample has been used in Ref.~\onlinecite{guandalini_nanolett2023} to quantify the importance of excitonic effects in the position and shape of the onset and $\pi$ plasmon in EEL spectra, and we approached the optical limit as close as we could reliably analyze the gap opening after subtraction of the zero-loss peak (ZLP).
Here, we move even closer to the optical limit, showing the relevance of low-momentum transfer kinematic effects, a broad topic relevant for all 2D materials which encompasses graphene physics.

We demonstrate that scattering geometry effects, encoded in the kinematic prefactor are responsible for a spectral reshaping akin to HREELS \cite{Palmer_prl1987, Palmer_surfsci1987, Li_surfsci2022}, 
need to be properly disentangled from the intrinsic dielectric response of the target in order to study the relative intensities of different EEL peaks and the low-dimensional effects on screening. 

\section{Results and discussion}

\subsection{2D kinematic factors}\label{sec:kin_prefactor}

\begin{figure}[h]
		\centering
		\includegraphics[width=0.38\linewidth]{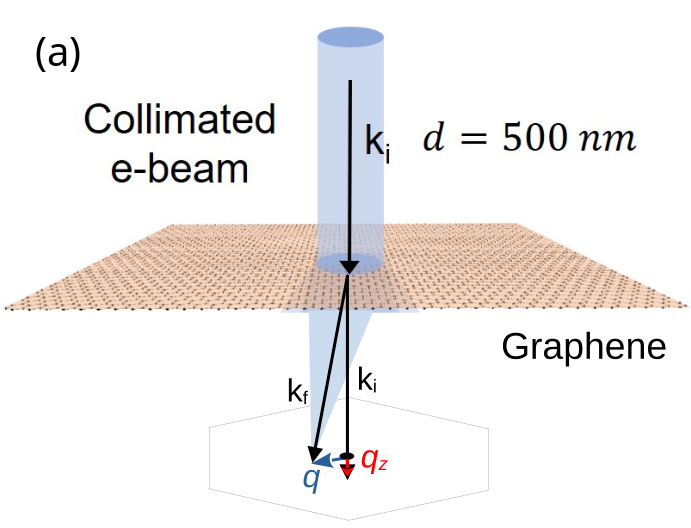}%
        \includegraphics[width=0.5\linewidth]{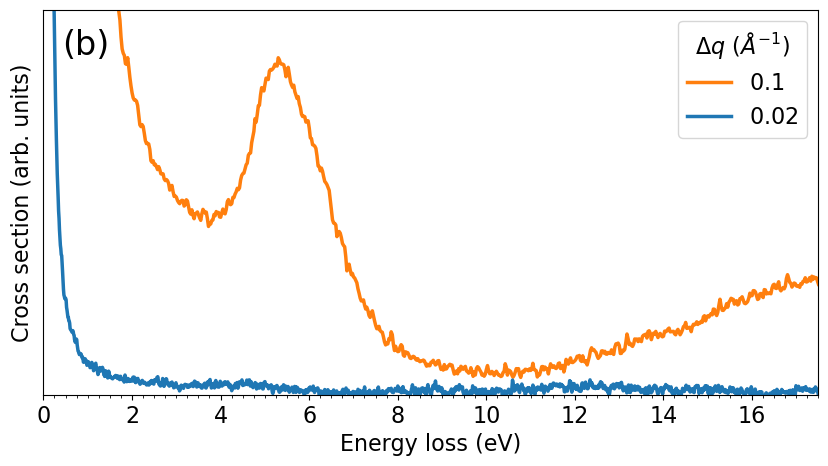}
		\caption{(a): parallel beam setup for momentum resolved EELS. The momentum transfer lost by the incoming electron along the direction parallel (perpendicular) to the beam, $\qz$ ($\qm$), is shown in red (blue). (b): EEL cross section at vanishing momentum transfer of graphite with standard $q$ resolution (orange) and of graphene with high resolution (blue).}
		\label{figIntro}
\end{figure}

In the (b) panel of Fig.~\ref{figIntro}, we show the optical limit of the EEL cross section of graphite with a momentum resolution of $\Delta q = 0.1$ \AA$^{-1}$ and of freestanding graphene at the higher momentum resolution ($\Delta q = 0.02$ \AA$^{-1}\approx \qz$).
We note the graphene cross section is completely quenched, as opposite to the case of graphite.

As we derive in the supporting information~\cite{supp-info}, the electron energy-loss cross section of a 2D system near the optical limit may be written as
\begin{equation}
\frac{d^2\crsec}{d\Omega d\hbar\omega}(|\qmv|\approx 0,\omega)
= \frac{4}{\pi a_0^2}\frac{\qm^2}{(\qm^2+\qz^2)^2}
\frac{\mathrm{Re}\left[ \tilde{\sigma}(\omega)\right]}{\omega}\ ,
\label{eq:cross_section_qlow}
\end{equation}
where $a_0$ is the Bohr radius
and $\tilde{\sigma}$ is the optical conductivity along the momentum transfer $\qmv$.
The EEL cross section near the optical limit contains the same information as the optical coefficients (see the supporting information~\cite{supp-info}). However, the
kinematic prefactor is $ I_{kin}\equiv \qm^2/(\qm^2+\qz^2)^2$, thus the cross section is quenched for  $\qm \le \qz$, even though the optical conductivity remains
finite.
Using energy and momentum conservation, $\qz$ can be expressed as\cite{Egerton_book,Roth_jes2014}:
\begin{eqnarray}
\qz(\qm,\omega) &=& k_i -\sqrt{k_i^2-\qm^2-2m\omega/\hbar}\label{eq:q_perp_full}\\
&\approx& \frac{m\omega}{\hbar k_i}+\frac{\qm^2}{2k_i} \,
\label{eq:q_perp}
\end{eqnarray}
where $k_i$ and $m$ are the wavenumber and mass of the incoming electron.
The approximation in the second-line holds as the energy of the beam electrons, $E_i \gg \hbar\omega$ and $k_i \gg \qm$.
The kinematic prefactor $I_{kin}$ encompasses, through $\qz$, a non trivial dependence of the EELS cross section on both the energy loss $\hbar\omega$ and the electron beam energy $E_i$, as given in Eqs.~\eqref{eq:q_perp_full}-\eqref{eq:q_perp}. In order to ease its deconvolution from the intrinsic response in momentum-resolved spectra, it is convenient to introduce a spectral reshape function defined as
\begin{equation}
\kfact(\qmv,\omega,E_i) \equiv\frac{\qm^4}{[\qm^2+\qz^2(\qm,\omega,E_i)]^2},
\label{eq:kin_factor}
\end{equation}
which connects the measured EELS cross section to the longitudinal dielectric response of the system.
Since EEL spectra are always expressed up to a multiplicative constant that takes into account experimental paramenters like exposure times, beam current etc., we define the spectral reshape function such that
it equals one in the limit $\qm \to 0$ and $\omega \to 0$. 
Being $A(\bm q,\omega,E_i)=q^2 I_{kin}$, our definition can be safely applied to the analysis of momentum-resolved spectra at fixed $\qm$, and the intrinsic response of the target can be obtained as the EEL spectrum divided by the reshape function.

\begin{figure}[h]
	\centering
	\includegraphics[width=0.7\linewidth]{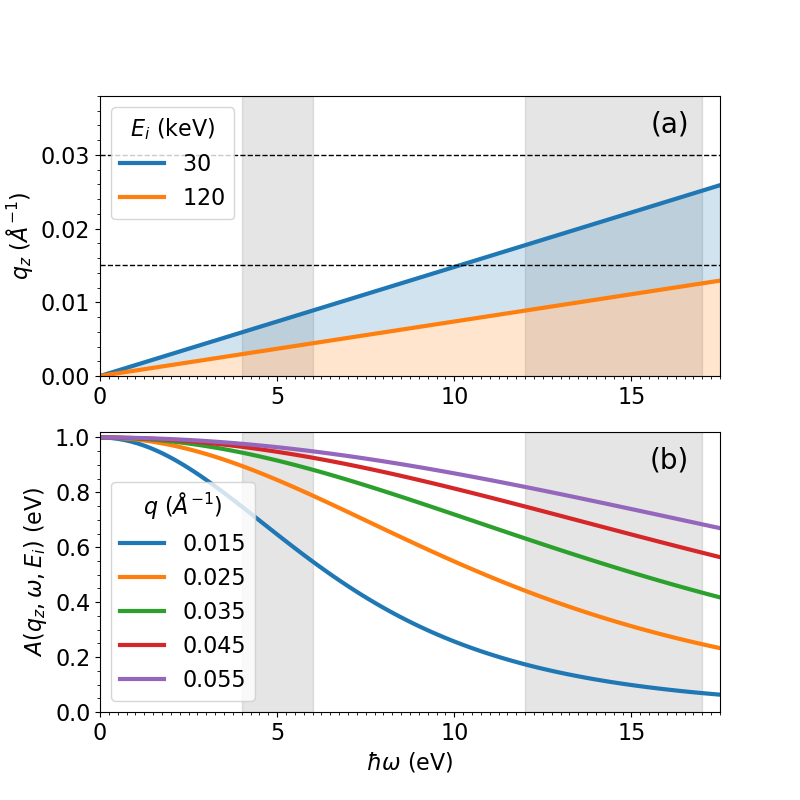}\\
	\caption{Energy loss dependence of $\qz$ (a) and the reshape factor $\kfact$ (b) for EEL spectra at different $\qm$ and incident energies $E_i$.
    $\qz$ is evaluated at $\qm = 0$, while $\kfact$ at $E_i = 30$ keV.
    Dashed horizontal lines in the (a) panel mark the lowest $\qm$ values shown in Fig.~\ref{figLowLossGK}.
    The $\pi$ and $\pi+\sigma$ plasmons of graphene lie in the gray spectral regions.}
	\label{figKinSuppression}
\end{figure}

We show in Fig.~\ref{figKinSuppression} concrete examples of $\qz$ and $\kfact$ dependencies on the energy loss $\hbar\omega$. 
In the (a) panel, we show $\qz$ at $\qm = 0$ for different incoming electron energies $E_i$. 
We verified numerically that $\qz$ is  independent on $\qm$ for the range considered ($\qm < 0.1$ \AA$^{-1}$, $\hbar\omega \approx$ eV). 
For 2D systems $q_z$ should be considered as a critical $q$ value under which the cross section is quenched due to 2D kinematic effects.
In the (b) panel of Fig.~\ref{figKinSuppression}, we show $\kfact$ at several $\qm$. The incoming electron energy $E_i$ is $30$ keV.  Each line shows the relative cross section drop due to the increase of $\qz$ as a function on $\hbar\omega$.
The drop in the signal is more pronounced at lower incoming energies $E_i$ (not shown).
At sufficiently small $\qm$, the signal is quenched in the $\hbar\omega$ range of the valence excitations.
The grey areas in Fig.~\ref{figKinSuppression} correspond to the energy ranges of graphene plasmons. The signal reshaping is very strong in the $\pi$ plasmon region at very low momentum transfer, while it becomes negligible at $\qm > 0.035$\AA$^{-1}$.
Instead the reshaping in the energy range of the $\pi+\sigma$ plasmon is approximately linear in $\omega$, and effective in the range of considered momentum transfer.
$\qz$ and $\kfact$ are system independent and thus apply to EEL spectra of any 2D system.

\subsection{Spectral reshaping at low momentum transfers}\label{sec_vanish_cr}

\begin{figure}[h]
		\centering
		\begin{minipage}{0.6\linewidth}
        \includegraphics[width=1\linewidth]{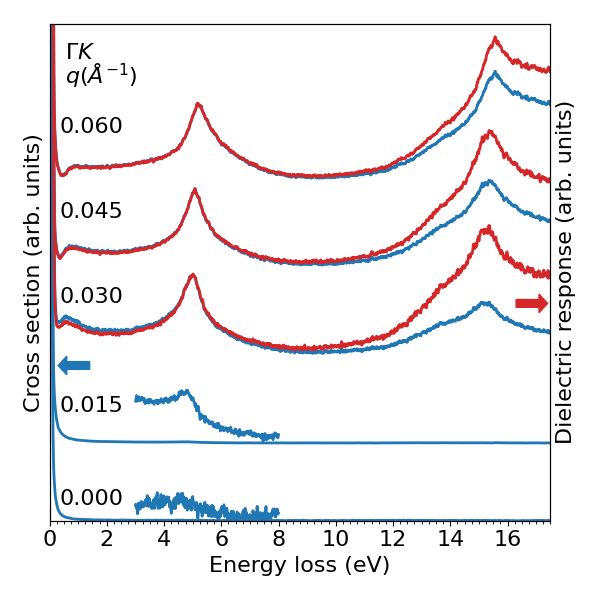}
        \end{minipage}\begin{minipage}{0.2\linewidth}
        \includegraphics[width=1\linewidth]{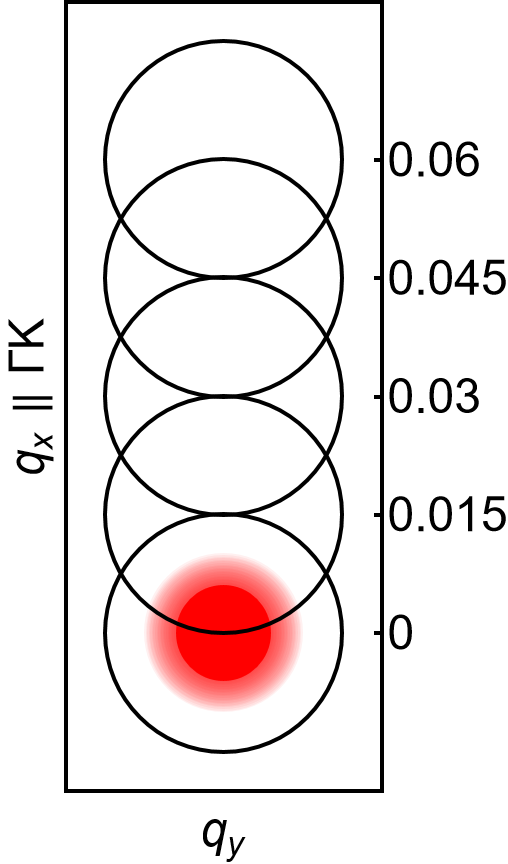}
        \end{minipage}
		\caption{
        Left: low energy EELS spectra of freestanding graphene at finite $ \qm$ along the $\Gamma$K direction and the nominal $\qm = 0$ spectrum. We show also the  $\times 40$ magnification of the two lower momentum transfer spectra in the range of the $\pi$ plasmon ($[3,8]$ eV). Red (blue) lines indicate EELS spectra (not) divided by the kinematic factor $\kfact$. Right:  sketch of the direct beam (red) and pin-hole aperture at different momentum transfers $\qm$.} 		
		\label{figLowLossGK}
\end{figure}

In Fig.~\ref{figLowLossGK}, we display with blue lines a set of EEL spectra measured at different low-momentum transfers $\qm$ and a beam energy of $30$~keV.
We emphasize the reduced width of the zero-loss peak (ZLP) attained with our experimental setup, where we adopt a pinhole-type aperture instead of a slit one.
We show in the right panel of Fig.~\ref{figLowLossGK} a sketch of the direct beam and the pinhole-type aperture at the different $\qmv$.
The two lowest pinhole positions correspond to the two spectra in the bottom of Fig.~\ref{figLowLossGK}. They are clearly different from earlier reports \cite{Kinyanjui2012,Wachsmuth2013,Nelson2014,Liou2015}, but also still affected by the direct beam.
The unprecedented observation of kinematically quenched plasmon peaks is attributed to the $\qm$ resolution of 0.02~\AA$^{-1}$.
The pinhole-type aperture does not only block the direct beam well below 0.1~\AA$^{-1}$, it is also instrumental for observing the quenched plasmon peaks near the optical limit.
In all the earlier reports \cite{Kinyanjui2012,Wachsmuth2013,Nelson2014,Liou2015} the loss spectra at the lowest $\qm$ were always the ones with the strongest plasmon peaks,
provided by the spurious inclusions of electrons with $q > \qz$, as $\Delta q > \qz$.

As the momentum transfer decreases down towards $0.030$ \AA$^{-1}$, the $\pi+\sigma$ plasmon loses spectral weight with respect to the $\pi$ plasmon.
At even lower momentum transfers, $\qm < 0.030$ \AA$^{-1}$, the spectral signatures of both the $\pi$ and $\pi+\sigma$ plasmons become comparable or smaller than the noise level, i.e., they are below the detection limit. To highlight the effect, we also plot the $\times 40$ zooms of the spectra for the two lowest momenta in Fig.~\ref{figLowLossGK}.
At $q \approx 0.015$ \AA$^{-1}$ a small trace of the $\pi$ plasmon can still be discerned from the spectral noise but the $\pi+\sigma$ plasmon (not shown) is already indiscernible.
As can be seen from the (a) panel of Fig.~\ref{figKinSuppression}, $q \approx 0.015$ \AA$^{-1}$ corresponds to the limit case where $\qm > \qz$ in the $\pi$ plasmon region and $\qm < \qz$ in the $\pi+\sigma$ plasmon region.
At $q \approx 0.000$ \AA$^{-1}$ the spectral noise is of the same order as the signal even in the energy range of the $\pi$ plasmon.
This is a clear signature of the cross section quenching due to kinematic effects at very low momentum transfer.
The signal quenching can be appreciated if only electrons with $\qm \le \qz$ are collected, thus for a $\qm$ resolution with $\Delta q$ at least on the order of $\qz$ as in our experimental setup. Otherwise, spurious inclusions of electrons with $\qm > \qz$ would dominate the cross section signal.
The faster quenching of the $\pi+\sigma$ spectral feature as opposed to the $\pi$ plasmon is consistent with the expected effect captured by the reshape function $\kfact$ shown in Fig.~\ref{figKinSuppression}. To further highlight the kinematic origin of the spectral reshaping in Fig.~\ref{figLowLossGK},
we also plot with red lines the spectra divided by the reshape function $\kfact$ from Eq.~\eqref{eq:kin_factor}. We excluded from this procedure the $\qm < 0.030$ \AA$^{-1}$ spectra because of the too high noise enhancement.
All such scaled spectra are normalized to have the same intensity as the originals at the $\pi$ plasmon peak.
The removal of kinematic effects increases the $\pi+\sigma$ intensity with respect to the $\pi$ peak intensity. 
The decrease of the intensity ratio between the $\pi+\sigma$ and $\pi$ peaks when approaching the optical limit is thus due to kinematic effects and it is not an intrinsic property of the dielectric response of graphene.

\subsection{Intrinsic dielectric response of Graphene}\label{sec_pi_pi+sigma}

\begin{figure}[htb]
		\centering
		\includegraphics[width=0.7\linewidth]{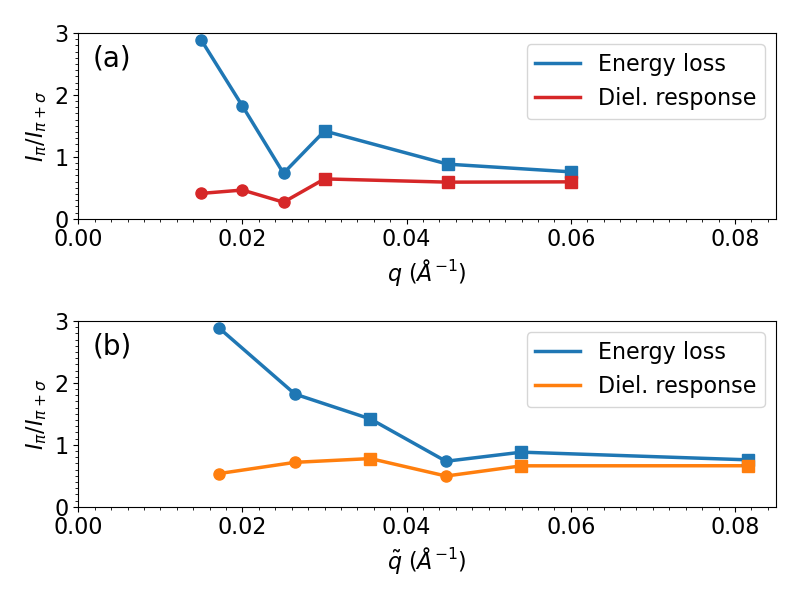}\\
		\caption{Ratio between the intensities of the $\pi$ and $\pi+\sigma$ plasmons of graphene as a function of the momentum transfer $\qm$. The upper panel uses the experimental $\qm$. The lower panel uses $\tilde{\qm}$ inferred from the isotropic $\pi$ plasmon dispersion \cite{Wachsmuth2014,guandalini_nanolett2023}. Squares (circles) are $\qm$ and $\tilde{\qm}$ along the $\Gamma$M ($\Gamma$K) direction. 
        }	
		\label{figratio}
\end{figure}

 In the  (a) panel of Fig.~\ref{figratio}, we show the ratio between the intensities of the $\pi$ and $\pi+\sigma$ plasmons as a function of momentum transfer $\qm$ for both the raw spectra and the dielectric response (obtained by dividing the spectra by $\kfact$).
Their spectral ratio $I_{\pi}/I_{\pi+\sigma}$ increases at lower momentum transfer due to the kinematic reshaping that lowers the intensity of the $\pi+\sigma$ plasmon more than that of the $\pi$ plasmon.
Once the signal is divided by the reshape function $\kfact$ and the intrinsic dielectric response is uncovered, the intensity ratio is instead approximately constant.
According to Eq. \eqref{eq:cross_section_qlow}, we deduce that in this regime the dielectric response is approximately converged to its optical value ($\qm = 0$), while kinematic effects are still very sensitive to small momentum transfer variations.

The finite accuracy of small momentum transfers is crucial for calculating $\kfact$. 
We address the experimental accuracy in the (b) panel of Fig.~\ref{figratio} by plotting the same intensity ratio with a momentum transfer $\tilde{\qm}$ calibrated on the well established isotropic linear $\pi$ plasmon dispersion~\cite{Wachsmuth2014,guandalini_nanolett2023}.
We note that a different momentum calibration does not just trivially change the $\tilde{\qm}$ values in the spectral intensity ratios, but it also alters the intensity ratios, as $\kfact$ also changes.
The qualitative picture is unchanged with the calibrated $\tilde{\qm}$, but the results are far less affected by experimental accuracy.
The differences between $\qm$ and $\tilde{\qm}$ are below $\approx 0.02$ \AA$^{-1}$, which is in line with the actual pixel size in the low magnification diffraction mode.
$\tilde{\qm}$ has the potential to become a valuable tools for further instrumental developments regarding small angle TEM-EELS. $\tilde{\qm}$ can be determined from one single EEL spectrum of graphene and then the experimental $\qm$ can be calibrated against $\tilde{\qm}$.

\section{Conclusions}\label{sec_conclusions}
We studied the EEL cross section of graphene near the optical limit.
We find the minimum observable $q$ is limited by the condition $q \ge q_z$, where $q_z$ is given by Eq.~\eqref{eq:q_perp}.
At lower momentum transfers, we observed a quenching of the EELS cross section in freestanding graphene. 
We have revisited the role of kinematic and scattering geometry effects in TEM-EELS cross section of low-dimensional, freestanding materials. Our analysis shows that the kinematic prefactor, entailing the well known dependence on both the energy loss and the energy of the beam electrons, is responsible for a quenching of the EELS cross section of atomically thin specimen when approaching the optical limit, albeit the intrinsic 2D optical conductivity remains finite.
High momentum resolution of TEM-EELS is instrumental to access the low-momentum transfer regime where such 2D kinematic effects emerge. In this respect, we argue that a wide collimated beam and a pinhole-type aperture in the diffraction plane are essential to achieve the best possible resolution required to go as close as possible to the optical limit. We emphasize however that our analysis is not limited to graphene, and it applies to any low-dimensional material of given thickness $d$ whenever the conditions $\qz d\ll 1$ and $\qm d\ll 1$ are satisfied. 

In order to ease the deconvolution of kinematic effects from the intrinsic 2D  response of the target, we introduced a reshape function for momentum-resolved EEL spectra. When applied to EEL spectra of graphene, the reshape function indicates that the relative intensity reduction of the $\pi+\sigma$ plasmon with respect to the $\pi$ plasmon is a pure kinematic effect, while the longitudinal dielectric response of the system approximately converges to its optical ($\qm\approx 0$) value. 

Since the high magnification of the diffraction plane does not allow to measure simultaneously Bragg spots usually needed to calibrate the momentum transfer, we benchmarked our accuracy in $q$ by comparing it against $\tilde{\qm}$ that was calibrated with the isotropic $\pi$ plasmon dispersion. 
The reshape function $\kfact$ from Eq.~\ref{eq:kin_factor} is -- via $\tilde{\qm}$ -- determined by the $\pi$ plasmon position alone and can be directly checked against the relative strengths of the $\pi$ and $\pi+\sigma$ plasmons in freestanding graphene. This may constitutes an independent and self consistent way to calibrate and benchmark the next generation of high angular resolution TEM-EELS instruments. 
  
\subsection*{Author Contributions}
A.G., P.B. and F.M. developed the theoretical framework.
Y.L. prepared the sample.
R.S. and Y.L. performed the experiments.
A.G. and C.K. analyzed the results.
All authors contributed to discussion of the results.
A.G., P.B., C.K., T.P. and F.M. wrote the manuscript.
\subsection*{Acknowlegements}
R.S. and K.S. acknowledges the support for JST-CREST (JPMJCR20B1, JPMJCR20B5, JPMJCR1993). This project has received funding from the European Research Council (ERC) under the European Union’s Horizon 2020 research and innovation programe (MORE-TEM ERC-SYN project, grant agreement No 951215).

\end{document}